\documentclass[twocolumn,showpacs,preprintnumbers,amsmath,amssymb]{revtex4}

\usepackage{graphicx}
\usepackage{dcolumn}
\usepackage{bm}

\begin{document}


\title{Time-dependent Nonequilibrium Thermodynamics:\\ A Master-equation Approach}

\author{Hao Ge}
\email{gehao@fudan.edu.cn}
 \affiliation{School of Mathematical Sciences
and Center for Computational Systems Biology, Fudan University,
Shanghai 200433, P.R. China}

\date{\today}

\begin{abstract}
Master equation could be applied to model various kinds of
biochemical systems. A general theory for its time-dependent
nonequilibrium thermodynamics is rigorously derived. We not only
introduce a concept of general internal energy, but also propose an
extension of the equilibrium state. Moreover, we find out an
explicit expression for the extended form of the Second Law, which
is phenomenologically stated as ``the conversion of work to excess
heat is irreversible''. The theory is carefully applied to several
thermodynamic processes, and its consistence with the classical
thermodynamics is clearly shown.

\end{abstract}


\maketitle

\section{Introduction}

Equilibrium thermodynamics emerges when Carnot proposed the first
theoretical treatise on mechanical work and efficiency in heat
engines in the early nineteenth century. However, one hundred fifty
years after its formulation, the second law of thermodynamics still
appears more as a program than a well defined theory, and all the
thermodynamic potentials could be well-defined only in equilibrium
states. This is one of the main reasons why the equilibrium
thermodynamics could hardly be applied to real biochemical systems,
because living cells must continually extract energy from their
surroundings in order to sustain the characteristic features of life
such as growth, cell division, intercellular communication, movement
and responsiveness to their environment.

The researches on irreversible systems far from equilibrium began
with the works by Haken \cite{Hak77,Hak83} about laser and
Prigogine, etc. \cite{GlPr,NP} about oscillations of chemical
reactions. Prigogine, I. and his collaborators provided explicit
expressions for entropy production in various situations, and
regarded a nonequilibrium steady state as a stationary open system
with positive entropy production rate \cite{NP}.

In 1998, Oono and Paniconi \cite{OP} proposed a framework of steady
state thermodynamics, and distinguished the steadily generated heat
which is generated even when the system remains in a single steady
state and the total heat. They called the former the ``housekeeping
heat'', which is equal to the entropy production in steady state and
may come from the chemical driven force in biochemical systems
\cite{QH05,QHBe05}. The key point of their work is that ``if we can
carefully remove the steadily produced heat due to housekeeping
dissipation, then the state should not be very different from
equilibrium''. Moreover, they also put forward a phenomenological
extended form of the Second Law: ``A process converting work into
excess heat is irreversible. And `reversibility' is modulo
house-keeping heat, which is produced anyway''.

On the other hand, it has been known for several decades that one
can use stochastic processes as mathematical tools to study
nonequilibrium states and steady cycle fluxes. In 1953, Onsager and
Machlup \cite{OnM1,OnM2} proposed the Onsager-Machlup principle,
which is actually a functional formula about the probability density
of a stochastic process close to equilibrium. T.L. Hill
\cite{Hi66,Hi77,Hi89,HC} and J. Schnakenberg \cite{Sc} successfully
constructed a general mesoscopic master-equation model for
biochemical systems and investigated its thermodynamic properties
far from equilibrium. Since then, a rather complete mathematical
theory for nonequilibrium steady states has been developed for
stochastic models \cite{JQQ}.

In recent years, a few interesting relations that describe the
statistical dynamics of driven systems even far from equilibrium
have been discovered, including the fluctuation theorems of sample
entropy production \cite{ECM,Kur,LeSp,GJ_JPA2007}, Jarzynski's
equality \cite{Jar1,Jar2,Jar3}, Hatano-Sasa equality \cite{HS}, etc.
The main purpose of Hatano and Sasa's work \cite{HS} is to derived
the first explicit expression for the extended form of the Second
Law of Thermodynamics, namely $T\triangle S\geq Q_{ex}$, where $S$
is the general entropy defined in their paper, and $Q_{ex}$ is the
excess heat.

In the present paper, we start with the definition of Gibbs entropy
\cite{Groot62} for stochastic processes, and accept the theory of
entropy production along a single stochastic trajectory
\cite{Seifert05}. Then, following the definitions of the three kinds
of heat \cite{HS,GJ_JSP2008}, it is found out the housekeeping heat
is always dissipated towards the environment, unless the
corresponding steady state is at equilibrium. Furthermore, the
dissipative work subtracting the excess heat along the trajectory
does not depend on the particular ``path'' taking through the
parameter space, namely, only depends upon the initial and final
states. Thus, there exactly exists a {\em ``general internal
energy''}, whose derivative is just the difference of the
dissipative work and excess heat.

Regarding the Second Law of Thermodynamics, it is shown that the
formulation here not only accords well to its former form, namely
the nonnegativity of entropy production, but also put forward
another explicit mathematical interpretation for its extended form,
which is phenomenologically stated as ``the conversion of work to
excess heat is irreversible'' \cite{OP,HS}. The extended form says
that the entropy production rate after subtracting the house-keeping
heat is still nonnegative, namely $T\cdot epr(t)-Q_{hk}(t)\geq 0$,
which consequently gives rise to the extended forms of the Clausius
inequality and Helmholtz/Gibbs free energy inequalities, and takes
Hatano-Sasa's work as a special example.

Similar result has been put forward in Langevin systems
\cite{Ge2009}, but the analysis here is much more comprehensive.

\section{Theory}

Now stochastic models are widely used in physics, chemistry, biology
and even in economics. The master-equation process discussed in this
article could be applied to model chemical reactions, which are of
special interest in biology, in relation with their coupling with
active transport across membrane \cite{Hi89,FMWT} and also recent
mechanisms of molecular motors \cite{KM}. Furthermore, in real
biochemical systems, the external parameters such as the
concentrations of external signal proteins always oscillate or
remarkably fluctuates, which stimulates the necessity for the
analysis of time-dependent processes.

In physics, master equation is a set of first-order differential
equations of the mesoscopic system:

\begin{equation}\label{MasterEq}
\frac{d}{dt}p_i(t)=\sum_{j=1}^N (q_{ji}(t)p_j(t)-q_{ij}p_i(t)),
\end{equation}
describing the dynamical evolution of a probability distribution
$p_i(t)$ over states $i=1,2,\cdots,N$. The quantity $q_{ij}(t)$ is
the transition density (probability per time) to state $j$ from
state $i$. It contains internal rate constants as well as external
conditions imposed by the coupling to the reservoir systems. The
basic properties of this model has been discussed previously
\cite{GJQ_JAP2006}.

One could take a stationary Markov chain, in which the probability
transition density $Q(t)$ is invariant, as the mathematical model of
the combination and transformation of biochemical polymers
\cite{Hi77,Hi89}. Each state of the Markov chain corresponds to a
mesoscopic state of polymers.

Let us also mention that the number $N$ need not to be finite, and
the system could also be regarded as the stochastic model of coupled
chemical reactions (chemical master equation) \cite{Sc,Mc,QE}.

\subsection{Essential Notations and Fundamental Relations}

Equilibrium thermodynamics is a generalization of mechanics by
introducing three new concepts: equilibrium states, internal energy
and entropy which related to disorder.

The concept of states is an ensemble of well-defined variables for
identifying any property of the system under study. Nowadays,
whether the most important quantities in thermodynamics, such as
temperature, volume, pressure, etc. could be generalized to
arbitrary nonequilibrium states or not is still unclear.
Fortunately, the situation is quite different when we study the
master-equation processes, in which the state variables are just
their transient distributions and specific time-dependent transition
laws.

However, the notation of internal energy seems unable to be directly
generalized to arbitrary state, and the First Law of Thermodynamics
remains unknown. In the present article, we will show that there
does exist an obvious way to generalize it to the time-dependent
Markov processes, enlightened by the mathematical equivalence of
Jarzynski and Hatano-Sasa equalities \cite{GQ_JMP2007,GJ_JSP2008}.

In the present article, we only investigate the isothermal system
with fixed temperature $T$. The role of temperature here is simply
to keep the environment constant, therefore we did not have to be so
careful about precise parameterization of the temperature. We simply
used the same temperature scale as the heat baths, and declared that
the temperature may be measured by a thermometer, just as Sasa
claimed in his recent review \cite{Sasa2006}.

\subsubsection{Gibbs entropy postulate and its generation}

Unlike the traditional approach of equilibrium thermodynamics, we
should start with the general definition of entropy, and then turn
back to consider the existence of internal energy. The common
definition of Gibbs entropy associated with any discrete probability
distribution $\{p_i\}$ is

$$S[\{p_i\}]=-k\sum_i p_i\log p_i,$$
where $k$ is the Boltzmann constant.

In statistical mechanics, it gives the entropy for a canonical
ensemble of a molecular system at constant temperature, and is a
generalization of Boltzmann's formula to a situation with nonuniform
probability distribution.

Denote the real distribution at time $t$ is $p(t)=\{p_i(t)\}$, and
we define the {\em general entropy} at time $t$ as
$S(t)=-k\sum_ip_i(t)\log p_i(t)$.

It is widely known that the entropy change $dS$ could be to
distinguished in  two terms \cite{NP,QH06,QH05}: the first, $d_eS$
is the transfer of entropy across the boundaries of the system, and
the second $d_iS$ is the entropy produced within the system.

Here, it is easy to derive that \cite{QH05,QH06}
\begin{equation}\label{EntropyEq}
\frac{dS(t)}{dt}=d_iS+d_eS=epr(t)-hdr(t),
\end{equation}
where
$$epr(t)=d_iS=\frac{1}{2}k\sum_{i,j}(p_i(t)q_{ij}(t)-p_j(t)q_{ji}(t))\log\frac{p_i(t)q_{ij}(t)}{p_j(t)q_{ji}(t)}$$
is just the instantaneous entropy production rate
\cite{GJQ_JAP2006}, and
$$hdr(t)=d_eS=\frac{1}{2}k\sum_{i,j}(p_i(t)q_{ij}(t)-p_j(t)q_{ji}(t))\log\frac{q_{ij}(t)}{q_{ji}(t)}$$
is due to the exchange of heat with the exterior, called the heat
dissipation rate.

\subsubsection{Entropy production along a stochastic trajectory}

A trajectory point of view will help us to realize the stochastic
nature of kinetics profoundly. Regarding its stochastic behavior,
any individual particle of the ensemble is governed entirely by the
transition density $Q(t)$ of the system, and ensemble properties
depend on their statistics.

We write the trajectory (sample path) of the Markov chain as $X(t)$.
Denote by $n_T$ the number of times that $X$ jumps in the time
interval $[0,T]$. Let $T_0=0$, $T_1=\inf\{t>0:X(t)\neq X(0)\}$,
$T_k=\inf\{t>T_{k-1}: X(t)\neq X(T_{k-1})\}$, and $T_{n_T+1}=T$ be
the jumping times.

Seifert \cite{Seifert05} defined the sample trajectory entropy along
the trajectory at time $t$ as $S(X(t),t)=-k\log p(X(t),t)$,
recalling $p(t)$ is the real distribution of the Markov chain
$\{X(t)\}$ at time $t$.

The equation for the motion of $S(X(t),t)$ becomes

$$\frac{d S(X(t),t)}{dt}=epr(X(t),t)-hdr(X(t),t),$$
where
\begin{eqnarray}
epr(X(t),t)&=&-k\frac{\partial \log p(X(t),t)}{\partial
t}\nonumber\\
&&-k\sum_{T_k}\delta_{t=T_k}\log\frac{p_{X(T_k)}(t)q_{X(T_k)X(T_{k-1})}(t)}{p_{X(T_{k-1})}(t)q_{X(T_{k-1})X(T_k)}(t)}.\nonumber
\end{eqnarray}
is the sample entropy production at time $t$, and
$$hdr(X(t),t)=-k\sum_{T_k}\delta_{t=T_k}\log\frac{q_{X(T_k)X(T_{k-1})}(t)}{q_{X(T_{k-1})X(T_k)}(t)}$$
is the sample heat dissipation in the medium.

The rationality after these identifications for the change rate of
entropy becomes clear when averaging over all the trajectories, then
one gets $epr(t)=\langle epr(X(t),t)\rangle$ and $hdr(t)=\langle
hdr(X(t),t)\rangle$.

It is also pointed out by Seifert \cite{Seifert05} that the
trajectory-dependent entropy of the particle could be measured
experimentally for a time-dependent protocol by first recording over
many trajectories the probability distribution $p_i(t)$; from which
the entropy of each trajectory can be inferred.

\subsubsection{Decomposition of heat conduction}

The idea of decomposing the total heat into a ``housekeeping'' part
and another ``excess'' part was put forward by Oono and Paniconi
\cite{OP}, and made explicit in Langevin systems by Hatano and Sasa
\cite{HS}.

The sample heat dissipation $hdr(X(t),t)$ could be regarded as the
total heat conduction $Q_{tot}(X(t),t)$ with the medium, i.e.
$$Q_{tot}(X(t),t)=T\cdot hdr(X(t),t),$$ and its ensemble
average $Q_{tot}(t)=T\cdot hdr(t)$. By convention, we take the sign
of heat to be positive when it flows from the system to the heat
bath.

For any fixed $t$, there is a steady distribution
$\pi(t)=\{\pi_i(t)\}$ corresponding to $Q(t)$ satisfying
$\pi(t)Q(t)=0$, which need not obey the detailed balance condition
$\pi_i(t)q_{ij}(t)=\pi_j(t)q_{ji}(t)$.

Then we could define the other two kinds of heat: the housekeeping
heat and excess heat along the trajectory
\cite{HS,GQ_JMP2007,GJ_JSP2008}:

$$Q_{hk}(X(t),t)=-kT\sum_{T_k}\delta_{t=T_k}\log\frac{\pi_{X(T_k)}(t)q_{X(T_k)X(T_{k-1})}(t)}{\pi_{X(T_{k-1})}q_{X(T_{k-1})X(T_k)}(t)},$$

$$Q_{ex}(X(t),t)=kT\sum_{T_k}\delta_{t=T_k}\log\frac{\pi_{X(T_k)}}{\pi_{X(T_{k-1})}},$$
and obviously $Q_{tot}(X(t),t)=Q_{hk}(X(t),t)+Q_{ex}(X(t),t)$.

After averaging over all the trajectories, one gets
$$Q_{ex}(t)=\frac{1}{2}kT\sum_{i,j}(p_i(t)q_{ij}(t)-p_j(t)q_{ji}(t))\log\frac{\pi_j(t)}{\pi_i(t)},$$

$$Q_{hk}(t)=\frac{1}{2}kT\sum_{i,j}(p_i(t)q_{ij}(t)-p_j(t)q_{ji}(t))\log\frac{\pi_i(t)q_{ij}(t)}{\pi_j(t)q_{ji}(t)},$$
and also $Q_{tot}(t)=Q_{ex}(t)+Q_{hk}(t)$.

More importantly, we found out that the housekeeping heat is always
nonnegative, which implies the nonequilibrium essence of the system:

\begin{eqnarray}
Q_{hk}(t)&=&kT\sum_{i,j}p_i(t)q_{ij}(t)\log\frac{\pi_i(t)q_{ij}(t)}{\pi_j(t)q_{ji}(t)}\nonumber\\
&\geq&-kT\sum_{i,j}p_i(t)q_{ij}(t)(\frac{\pi_j(t)q_{ji}(t)}{\pi_i(t)q_{ij}(t)}-1)\nonumber\\
&=&-kT\sum_{i}\frac{p_i(t)}{\pi_i(t)}\sum_j\pi_j(t)q_{ji}(t)+kT\sum_{i,j}p_i(t)q_{ij}(t)\nonumber\\
&=&0,\nonumber
\end{eqnarray}
by making use of a simple inequality $\log x\leq x-1$ for $x>0$, and
the identity $\sum_{j}q_{ij}(t)\equiv 0$.

For equilibrium system, $Q_{ex}$ reduces to the total heat
$Q_{tot}$, because in this case $Q_{hk}\equiv 0$ due to
$\pi_i(t)q_{ij}(t)=\pi_j(t)q_{ji}(t)$. And in time-independent
steady state, $Q_{ex}(t)\equiv 0$, and hence the housekeeping heat
$Q_{hk}$ equals the work done by the external driven force, which is
all dissipated \cite{QH06,QHBe05}.

However, the situation is quite different for the time-dependent
nonequilibrium system, in which the housekeeping heat still comes
from the work done by some external driven force but where does the
excess heat come from?We will show that its origin is just the
change of a thermodynamic quantity called {\em general internal
energy}.

\subsubsection{Dissipative work and general internal energy}

Notice that the heat conduction only occurs at the specific time
when the sample trajectory jumps, but which manner of energy
exchange would happen during the time interval as the trajectory
stays in the same state? It is surely due to the variation of the
time-dependent transition laws from the external perturbation acting
on the system. With the classic quantum system in mind, we realize
that it is just the ``dissipative work'' called by Jarzynski and
Crooks \cite{Jar1,Jar2,Jar3,Cro1,Cro2,Cro3}, and rigorously
formulated by Min Qian and the author \cite{GQ_JMP2007}.

The dissipative work along the trajectory is defined as
\cite{HS,GQ_JMP2007,HuS}
$$W(X(t),t)=-kT\frac{\partial\log\pi_{i}(t)}{\partial t}|_{i=X(t)},$$
with ensemble average
$W(t)=-kT\sum_ip_i(t)\frac{d\log\pi_i(t)}{dt}$.

If the system satisfies the detailed balance conditions for all
time, i.e. $\pi_i(t)q_{ij}(t)=\pi_j(t)q_{ji}(t)$, then the
traditional concept of internal energy exists and both of the excess
heat and dissipative work contribute to its change, which is the
First Law undoubtedly \cite{GQ_JMP2007}. Therefore, we believe that
the situation will not be essentially different even if detailed
balance conditions fails.

Further, we find out that the dissipative work subtracting the
excess heat along the trajectory does not depend on the particular
``path'' taking through the parameter space, namely, only depends
upon the initial and final states, i.e.
$$W(X(t),t)-Q_{ex}(X(t),t)=dU(X(t),t),$$
where $U(X(t),t)=-kT\log\pi_{X(t)}(t)$ could be regarded as the
internal energy of state $X(t)$.

Thus, there exactly exists a {\em ``general internal energy''}
$U(t)=\langle U(X(t),t)\rangle=-kT\sum_{i}p_i(t)\log\pi_i(t)$, whose
derivative is just the difference of the dissipative work and excess
heat, i.e.
\begin{equation}\label{Firstlaw}
\frac{dU(t)}{dt}=-Q_{ex}(t)+W(t).
\end{equation}
It is just the ordinary internal energy for the equilibrium
canonical ensemble according to the Maxwell-Boltzmann's law. Hence
here Eq.(\ref{Firstlaw}) is just the generalized First Law of
thermodynamics.

\subsubsection{General free energy and the concept of free heat}

Based on the elementary definition of free energy in equilibrium
thermodynamics $F=U-TS$, here we could define a {\em general free
energy} in the same way:
$$F(t)=U(t)-TS(t)=kT\sum_ip_i(t)\log\frac{p_i(t)}{\pi_i(t)},$$
which is just the relative entropy of the distribution $\{p_i(t)\}$
with respect to another one $\{\pi_i(t)\}$ from the mathematical
point of view \cite{QH01b}.

For equilibrium system, it is just the Gibbs free energy in a
spontaneously occurring chemical reaction at constant pressure $p$
and temperature $T$, and also the Helmholtz free energy for systems
at constant $V$ and $T$ \cite{Ross2008}. Its change gives the
maximum work, other than pV work. Therefore, it is called a ``hybrid
free energy'' by Ross, J. \cite{Ross2008}.

More important, Schnakenberg \cite{Sc} has shown that it is just the
Lyapunov function as well as Prigogine-Glansdorff criterion
certificating the thermodynamic stability for the steady state of
the time-independent master equation system. This will be revisited
in Sec. \ref{Sec_secondlaw}.

Due to the Jensen's equality for the convex function $-\log x$, we
have $F(t)\geq 0$, and the equality holds if and only if
$p_i(t)=\pi_i(t)$, for each state $i$.

On the other hand,
\begin{eqnarray}\label{FreeEq}
\frac{dF(t)}{dt}&=&\frac{dU(t)}{dt}-T\frac{dS(t)}{dt}\nonumber\\
&=&W(t)-(T\cdot epr(t)-Q_{hk}(t)),
\end{eqnarray}

Here we introduce a new concept named {\em Free heat} $Q_f(t)=T\cdot
epr(t)-Q_{hk}(t)$ identifying the free energy change in the form of
heat, i.e.
$$\frac{dF(t)}{dt}=W(t)-Q_f(t).$$
This concept will play the central role in the extended form of the
Second Law of Thermodynamics below.

\subsection{Energy Balance and External Driven Force}

The First Law of Thermodynamics is essentially an extension of the
principle of the conservation of energy to include systems in which
there is flow of heat. But the situation becomes more complicated in
nonequilibrium case, since there may exists some external driven
force which also pumps energy into the system but does not
contribute to the change of internal energy \cite{QHBe05}.

Denote the work done by the external driven force as $Edf(t)$, and
we here try to figure out its relationship with other quantities
defined in the previous sections.

According to energy balance, one has
$$W(t)+Edf(t)=\frac{dU(t)}{dt}+T\cdot hdr(t).$$

On the other hand, we have already known that

$$\frac{dU(t)}{dt}=-Q_{ex}(t)+W(t)=-Q_{tot}(t)+Q_{hk}(t)+W(t),$$
and also $Q_{tot}(t)=T\cdot hdr(t)$.

Therefore, it yields $Edf(t)=Q_{hk}(t)$, which is only known to be
valid in steady state before.

Now we understand that there exist totally two kinds of external
works done on the system, one is the dissipative work $W(t)$ and the
other $Edf(t)$ from the external driven force. They result in the
change of general internal energy and the heat dissipation
respectively.

\subsection{Second Law of Thermodynamics}
\label{Sec_secondlaw}

Traditional Second Law of Thermodynamics has two kinds of statements
\cite{Adkins83}: The Kelvin-Planck statement ``No process is
possible whose sole result is the complete conversion of heat into
work'' and the Clausius statement ``No process is possible whose
sole result is the transfer of heat from a colder to a hotter
body''.

Now, a central problem arises: how does these thermodynamic laws
apply to such a nonequilibrium time-dependent process? First, we
provide a rigorous quantitative approach to the Clausius inequality
and Helmhotz/Gibbs free energy inequalities based on the
nonnegativity of entropy production rate for nonequilibrium
time-dependent processes.

Then a new concept ``instantaneous reversible process''
\cite{GJQ_JAP2006} with zero entropy production rate natually
emerges, which corresponds to the ideal reversible process involved
in the classic theory of equilibrium thermodynamics, and it will
imply that there does not exist any real reversible process
connecting two different equilibrium states.

Furthermore, an extended quantitative form of Second Law of
Thermodynamics will be developed built on the nonnegativity of the
new concept {\em ``free heat''}, which only appears during
time-dependent processes and accords well to the phenomenological
statement by Oono and Panicini \cite{OP}.

\subsubsection{Former quantitative forms of the Second Law}

Although all the thermodynamic quantities in the previous sections
could be defined along the sample trajectory, the Clausius
inequality and many other thermodynamic constrains related to the
Second Law should be interpreted statistically through ensemble
average.

Notice that every term in the expression of the entropy production
rate, i.e.
$$epr(t)=\frac{1}{2}k\sum_{i,j}(p_i(t)q_{ij}(t)-p_j(t)q_{ji}(t))\log\frac{p_i(t)q_{ij}(t)}{p_j(t)q_{ji}(t)}$$
is nonnegative, and the equality holds if and only if
$p_i(t)q_{ij}(t)=p_j(t)q_{ji}(t)$ for each pair of states $i$ and
$j$.

Then according to Eqs. (\ref{EntropyEq}) and (\ref{FreeEq}), we
derived several inequalities of the differential forms:

\begin{subequations}
\label{Secondlaw}
\begin{eqnarray}
T\frac{dS(t)}{dt}+Q_{tot}(t)&=&T\cdot epr(t)\geq 0,\label{Secondlaw1}\\
\frac{dF(t)}{dt}-W(t)-Q_{hk}(t)&=&-T\cdot epr(t)\leq
0.\label{Secondlaw2}
\end{eqnarray}
\end{subequations}

Eq. (\ref{Secondlaw1}) is just the well-known Clausius inequality
($dS\geq -\frac{Q_{tot}}{T}$), which is rectified to obtain
expressions for the entropy produced ($dS$) as the result of heat
exchanges ($Q_{tot}$). And Eq. (\ref{Secondlaw2}) is a general
version of the free energy inequality for the amount of work
performed on the system, since the work values must then be
consistent with the Kelvin-Planck statement \cite{Finn93} and
forbids the systematic conversion of heat to work.

More precise, the quantity $Q_{hk}(t)$ in Eq. (\ref{Secondlaw2})
vanishes when the detailed balance condition
($\pi_i(t)q_{ij}(t)=\pi_j(t)q_{ji}(t)$) holds£¬ and then it returns
back to the traditional Helmhotz or Gibbs free energy inequalities
of equilibrium thermodynamics depending on whether it is a NVT or
NPT system \cite{BQ2008}. In this case, $-dF\geq -W$, which implies
the decrease of free energy gives the maximum dissipative work done
upon the external environment.

Their corresponding integral forms are
\begin{subequations}
\label{Secondlawint}
\begin{eqnarray}
T\triangle S+\int Q_{tot}(t)dt&\geq& 0,\label{Secondlawint1}\\
\triangle F-\int W(t)dt-\int Q_{hk}(t)dt&\leq&
0.\label{Secondlawint2}
\end{eqnarray}
\end{subequations}

\subsubsection{Reversible process}

Reversible process that is beneficial for thermodynamic studies, is
defined as a process that, once having taken place, can be reversed,
and in so doing leaves no change in either the system or
surroundings. However, there still leaves a fundamental question:
``what is precisely the reversible process that connects two
different equilibrium states?'' Since it is well-known that ``only
irreversible processes contribute to entropy production'', what we
need here is only to check the condition for which the entropy
production vanishes. Then we find out that it is just equivalent to
the concept ``instantaneous reversibility'' \cite{GJQ_JAP2006},
which is a natural generalization of equilibrium state.

For the time-dependent process discussed in the present article, let
$epr(t)=0$ for each time $t$, then we find that all the steady
distributions $\{\pi(t)\}$ must be independent with time $t$ (i.e.
$\pi(t)\equiv$ some fixed distribution $\pi$) and the detailed
balance condition holds, i.e. $\pi_iq_{ij}(t)=\pi_jq_{ji}(t)$.
Therefore, during this process, the transient state at each time $t$
is just the real equilibrium state corresponding to the transition
law $Q(t)$. But unfortunately, these equilibrium states are all
essentially the same as the initial one, only with different time
scales, i.e. $Q(t)=f(t)\cdot Q(0)$ for some function $f$. In other
words, there is no real reversible process between two different
equilibrium states, which confirms the well-known belief in
equilibrium thermodynamics. This reversible process is called the
``instantaneous reversible process'' \cite{GJQ_JAP2006}, which could
be realized as a generalization of the concept of equilibrium state.

Note that, all the fundamental equations in classic equilibrium
thermodynamics, such as $dU=TdS-pdV$, requires that initial and
final equilibrium states be defined and that there is some
reversible path between them \cite{Adkins83}. But unfortunately it
is not ture, and this is just why equilibrium thermodynamics could
not be directly generalized to the far-from-equilibrium case.

\subsubsection{Extended quantitative forms of the Second Law}

Oono and Paniconi \cite{OP} put forward a phenomenological extended
form of the Second Law to nonequilibrium steady states, which claims
the irreversibility of the process converting work into excess heat
when modulo house-keeping heat. After that, a quantitative
expression in Langevin systems was given for the transition
processes between steady states  by Hatano and Sasa \cite{HS}.

Here we give a rather different but much more general derivation,
only need to notice that the free heat $Q_f(t)=epr(t)-TQ_{hk}(t)\geq
0$ again due to the simple inequality $\log x\leq x-1$ for $x>0$:
\begin{eqnarray}
Q_f(t)&=&T\cdot epr(t)-Q_{hk}(t)\nonumber\\
&=&\sum_{ij}p_i(t)q_{ij}(t)\log\frac{p_i(t)\pi_j(t)}{p_j(t)\pi_i(t)}\nonumber\\
&\geq &\sum_{ij}p_i(t)q_{ij}(t)(\frac{p_j(t)\pi_i(t)}{p_i(t)\pi_j(t)}-1)\nonumber\\
&=&\sum_{j}\frac{p_j(t)}{\pi_j(t)}\sum_{i}\pi_i(t)q_{ij}(t)-\sum_{i}p_i(t)\sum_{j}q_{ij}(t)\nonumber\\
&=&0,
\end{eqnarray}

Then according to Eqs. (\ref{EntropyEq}) and (\ref{FreeEq}), we have
another group of  thermodynamic inequalities in the differential
forms:

\begin{subequations}
\label{Secondlaw_ex}
\begin{eqnarray}
T\frac{dS(t)}{dt}+Q_{ex}(t)&=&T\cdot epr(t)-Q_{hk}(t)\geq 0,\label{Secondlaw_ex1}\\
\frac{dF(t)}{dt}-W(t)&=&-Q_f(t)=-T\cdot epr(t)+Q_{hk}(t)\leq
0.\nonumber\\
\label{Secondlaw_ex2}
\end{eqnarray}
\end{subequations}
followed by their corresponding integral forms
\begin{subequations}
\label{Secondlawint_ex}
\begin{eqnarray}
T\triangle S+\int Q_{ex}(t)dt&\geq& 0,\label{Secondlawint_ex1}\\
\triangle F-\int W(t)dt&\leq& 0.\label{Secondlawint_ex2}
\end{eqnarray}
\end{subequations}

Eq. (\ref{Secondlawint_ex1}) is the extended form of Clausius
inequality during any nonequilibrium time-dependent process, whose
special case is included in Hatano and Sasa's work \cite{HS}. And
Eq. (\ref{Secondlawint_ex2}) is a different general form of free
energy inequality. It implies the dissipative work value must be
consistent with the Oono-Paniconi statement of the extended Second
Law of thermodynamics \cite{OP}, which forbids the systematic
conversion of excess heat to work.

For equilibrium case, $Q_{ex}=Q_{tot}$, then they both return back
to Eq. (\ref{Secondlaw1}) actually. And then if in steady state,
then $Q_f(t)\equiv 0$, and this form of the Second Law completely
disappears.

\section{Applied to classic thermodynamic processes}

\subsection{Time-independent case}

\subsubsection{Equilibrium states and nonequilibrium steady states}

Here we consider the Markov chain with transition density matrix
$Q=\{q_{ij}\}$ and stationary distribution $\pi=\{\pi_i\}$
satisfying $\pi Q=0$.

The first step is to distinguish equilibrium and steady
nonequilibrium states. As is well known, thermodynamic equilibrium
is in general maintained through detailed balance, which was already
known by Boltzmann; and the reversibility of a time-independent
stochastic process was first introduced by Kolmogorov. It has been
rigorously proved by Min Qian, Minping Qian and their collaborators
that the mathematical essence of these two concepts actually turns
out to be the same, and the major characteristic of a nonequilibrium
steady state is shown to be the positivity of enropy production rate
and the appearance of circulations \cite{JQQ}.

Note that in the stationary time-independent case, the general
internal energy, entropy, and free energy are all independent of
time, and both of the dissipative work and excess heat vanish. That
may be the reason why these concepts have not been found previously,
when only dealing with steady states.

For steady states, we have the important relations
$$T\cdot hdr=T\cdot epr=Q_{hk}=Q_{tot}\geq 0,$$
and the equality holds if and only if at equilibrium states.

Moreover, the energy balance tells us that the external driven force
$Edf=T\cdot hdr$, which implies the external energy pumped in all
changes into heat dissipation \cite{QH05,QH06}. And it is also
consistent with the truth that there is no external energy source
for equilibrium states in which $hdr\equiv 0$.

\subsubsection{Relaxation process towards equilibrium states}

Each undriven system will be found to reach a state where no further
change takes place and it is then said to have come to thermodynamic
equilibrium. In general, the relaxation to thermodynamic equilibrium
will involve both thermal and work-like interactions with the
surroundings, and then which will occur during this relaxion
process?

First, the time-independent property implies the vanishing of
dissipative work, i.e. $W(t)\equiv 0$, but the excess heat
$Q_{ex}(t)$ may not vanish. Second, since the final state of the
relaxation process is at equilibrium, so the external driven force
$Edf(t)$ disappears. Consequently, $Q_{hk}(t)=Edf(t)\equiv 0$, and
$Q_{tot}(t)$ reduces to $Q_{ex}(t)$.

However, it will still dissipate heat and have positive entropy
production rate unless it has already arrived at the final
equilibrium state. To be more precise,  we have
$hdr(t)=\frac{Q_{tot}(t)}{T}=\frac{Q_{ex}(t)}{T}$, therefore,
\begin{equation}
\frac{dU(t)}{dt}=-Q_{ex}(t)=-T\cdot hdr(t),
\end{equation}
which is just the First Law of Thermodynamics.

Furthermore,
\begin{equation}
\frac{dF(t)}{dt}=W(t)-T\cdot epr(t)+Q_{hk}(t)=-T\cdot epr(t)\leq 0,
\end{equation}
hence $F(t)$ is just the Lyapunov function while relaxing towards
equilibrium states \cite{Sc}. Until now we know that this Lyapunov
function is just a directly corollary of the Second Law.

\subsubsection{Relaxation process towards steady state}

The relaxation process towards steady states has been extensively
discussed by Glansdorff and Prigogine \cite{GlPr,NP}, and then by
Schnakenberg for the master-equation systems \cite{Sc}.

Comparing with the previous discussion on the relaxation process
towards equilibrium state, we will show that the sole difference
here only stays within the housekeeping heat provided by some
specific external driven force.

Here, the First Law of Thermodynamics also becomes
\begin{equation}
\frac{dU(t)}{dt}=-Q_{ex}(t),\nonumber
\end{equation}
and the energy balance reads
\begin{equation}
\frac{dU(t)}{dt}+T\cdot hdr(t)=Edf(t).\nonumber
\end{equation}

Combined with $Edf(t)=Q_{hk}(t)$, it is found that the whole heat
dissipation is from two sources: one is the excess heat contributing
to the change of general internal energy; the other is the
housekeeping heat caused by the external driven force.

The extended form of the Second Law now reads
\begin{equation}
\frac{dF(t)}{dt}=-T\cdot epr(t)+Q_{hk}(t)=-Q_f(t)\leq 0,\nonumber
\end{equation}
thus $F(t)$ serves as a Lyapunov function for this relaxation
process \cite{Sc}, which actually has a solid thermodynamic basis
now.

\subsection{Time-dependent case}

Time-dependent processes are causing more and more interests from
physicists nowadays \cite{Zuba74,Jar2008,HS,Seifert05}, and it will
uncover many important thermodynamic properties that originally
hidden behind the stationary time-independent case.

\subsubsection{Cyclic process}

In equilibrium thermodynamics, a thermodynamic cycle is a series of
thermodynamic processes which returns a system to its initial state.
As a conclusion of cyclic process, all the state variables should
have the same value as they had at the beginning. Thus $\triangle
U=\triangle S=\triangle F=0$.

But variables such as heat and work are not zero over a cycle, but
rather are process dependent. The First Law of Thermodynamics
dictates that the net heat input is equal to the net work output
over any cycle, i.e. $\int W(t)dt=\int Q_{ex}(t)dt$.

Hence in this case, the former form of the Second Law
(\ref{Secondlawint}) gives

\begin{subequations}
\label{CyclicSecondlaw}
\begin{eqnarray}
\int Q_{tot}(t)dt=T\cdot \int epr(t)dt&\geq& 0,\label{CyclicSecondlaw1}\\
\int W(t)dt&\geq& -\int Q_{hk}dt.\label{CyclicSecondlaw2}
\end{eqnarray}
\end{subequations}

If one rewrite Eq. (\ref{CyclicSecondlaw2}) as $\int
(W(t)+Edf(t))dt=\int Q_{tot}(t)dt\geq 0$, then it is just the
familiar statement of traditional Second Law of Thermodynamics ``the
conversion from work to total heat is irreversible''.

Moreover, the extended form (\ref{Secondlawint_ex}) says

\begin{subequations}
\label{CyclicSecondlaw_ex}
\begin{eqnarray}
\int Q_{ex}(t)dt=\int Q_f(t)dt&\geq& 0,\label{CyclicSecondlaw1_ex}\\
\int W(t)dt=\int Q_{ex}(t)dt&\geq& 0,\label{CyclicSecondlaw2_ex}
\end{eqnarray}
\end{subequations}
confirming the claim that ``the conversion from work to excess heat
is irreversible'' \cite{OP}. In other words, during a cyclic
process, not only the total heat but also the excess heat could only
be from the system into the heat bath rather than follow the
opposite direction.

\subsubsection{Transitions between equilibrium states}

Jarzynski provided an expression for the equilibrium free energy
difference between two configurations of a system, in terms of an
ensemble of finite-time measurements of the work performed through
switching from one configuration to the other (See a recent review
\cite{Jar2008} and references in).

In the stochastic-process approach
\cite{Jar2,GQ_JMP2007,GJ_JSP2008}, we consider the transition
between two equilibrium states realized in the time interval $[0,T]$
where the detailed balance condition is satisfied. In other words,
there is no external driven force, i.e. $Edf(t)=Q_{hk}(t)\equiv 0$.

Then the energy balance becomes
$$W(t)=\frac{dU(t)}{dt}+T\cdot hdr(t),$$
thus one could realize that it is not very reasonable to call the
quantity $W(t)$ as ``dissipative'', since part of it would be stored
in the general internal energy rather than really convert into heat.

Notice that when one finishes the task of driving the process from
time $0$ to time $T$ through modulating the time-dependent
transition density matrix $Q(t)$ (denoted as process $1$), the
system has not reached the final equilibrium state yet. Then we
should wait until it really arrives (denoted as process $2$), which
is just the ``relaxation process towards equilibrium state'' with
the fixed transition density $Q(T)$ described in the previous
subsections.

According to Eq. (\ref{Secondlawint2}), we have

$\triangle F_1\leq \int_1 W(t)dt$, and $\triangle F_2\leq \int_2
W(t)dt$ for process $1$ and $2$ respectively.

Finally, since $\int_2 W(t)=0$, and $\triangle F=\triangle
F_1+\triangle F_2$ is just the free energy difference between the
initial and final equilibrium states, we conclude $\triangle F\leq
\int_1 W(t)dt$. It is just why we could neglect process 2 when
applying Jarzynski's work relation in experiments.

Note that the situation is quite different when regarding
Hitano-Sasa's equality for transitions between steady states
\cite{HS}. It has not been explicitly pointed out in previous works
 \cite{HS,GQ_JMP2007,GJ_JSP2008,Jar2008}. See below for details.

\subsubsection{Transitions between steady states}

The main contribution of Hatano and Sasa \cite{HS,Sasa2006} is to
give the first explicit expression for the extended form of the
Second Law put forward by Oono and Paniconi \cite{OP}, during the
transition process between steady states.

They also found out a general definition of entropy enlighten from
the idea that any proper formulation of steady state thermodynamics
should reduce to equilibrium thermodynamics in the appropriate
limit, so that $Q_{ex}$ should correspond to the change of a
generalized entropy S.

However, what they defined is just the general internal energy
consistent with the First Law of Thermodynamics rather than the
general Gibbs entropy in the present article and also in Seifert's
recent work \cite{Seifert05}. Note that the two quantities are
always different except for steady states, hence the extend form of
Second Law of thermodynamics during the course of transition between
two steady states derived by Hatano and Sasa is not flawed.

Similar to the preceding transition process between two equilibrium
states, when one finishes driving the process from time $0$ to $T$
(also denoted as process $1$) through varying the time-dependent
transition density $Q(t)$ , the system has not reached the final
steady state yet, and we should wait until it arrives (denoted as
process $2$), which is just the ``relaxation process towards steady
state'' where the transition density is fixed at $Q(T)$.

According to Eq. (\ref{Secondlawint_ex}), we get

$\int_1 Q_{ex}(t)dt\geq -\triangle S_1$, and $\int_2 Q_{ex}(t)dt\geq
-\triangle S_2$ for process 1 and 2 respectively.

Finally, since $\triangle U=\triangle S=\triangle S_1+\triangle S_2$
representing the energy (entropy) difference between the initial and
final steady states, we derive that $$\int Q_{ex}(t)dt=\int_1
Q_{ex}(t)dt+\int_2 Q_{ex}(t)dt\geq -\triangle S.$$ Hatano and Sasa
\cite{HS} concluded that the equality held for an infinitely slow
operation in which the system is in a steady state at each time
during a transition(``slow process'').

It is indispensable to emphasize that we could not neglect process 2
this time when applying Hitano-Sasa's identity, because here $\int_2
Q_{ex}(t)dt$ may not be zero, which implies that the relaxation
process towards the final steady state will also contribute to the
heat dissipation. This critique has already be pointed out by Cohen
and Mauzerall \cite{Cohen04}, but unfortunately what they criticized
is the Jarzynski's equality rather than Hatano-Sasa equality. In
real experiment, this relaxation process may be rapid enough and
could somehow be omitted.

\subsection{Summary}

The essential difference between these typical processes relies
mainly on the signs of the three key thermodynamic quantities: the
housekeeping heat, entropy production and free heat. And also the
dissipative work would disappear for time-independent processes. See
Tab. \ref{tab1} for details.

\begin{table*}
\caption{The signs of important thermodynamic quantities. Process
$1$: Equilibrium state; Process $2$: Nonequilibrium steady state;
Process $3$: Relaxation process towards equilibrium state; Process
$4$: Relaxation process towards nonequilibrium steady state; Process
$5$: Transition between equilibrium states; Process $6$: Transition
between steady states. ``0'' for zero, ``+'' for positive, ``-'' for
negative, and ``?'' for uncertainty.} \label{tab1}
\begin{ruledtabular}
\begin{tabular}{ccccccccccccc}
&Process 1&Process 2&Process 3&Process 4&Process 5&Process 6\\
\hline
Dissipative work&0&0&0&0&?&?\\
Excess heat&0&0&?&?&?&?\\
Housekeeping heat&0&+&0&+&0&+\\
Entropy production&0&+&+&+&+&+\\
Free heat&0&0&+&+&+&+
\end{tabular}
\end{ruledtabular}
\end{table*}

\section{Discussion}

It is the main thesis of this article that we are only at the
beginning of a new development of theoretical chemistry and physics
in which thermodynamic concepts may play an even more basic role.
``In any case, the number of thermodynamic or macroscopic variables
is much less than the large number of the microscopic degrees of
freedom. Hence, the transition from a microscopic to a macroscopic
description involves a drastic reduction of the information about
the system.''\cite{Sc1}

However, the study of thermodynamics before is largely confined to
equilibrium states. Although the field of ``nonequilibrium
thermodynamics'' has successfully extended the 19th century concepts
of equilibrium thermodynamics to the systems that are close to, or
near equilibrium, the understanding of far-from-equilibrium systems
is still poor.

To investigate these points, stochastic thermodynamics has stepped
much further than other approaches during the last two decades
\cite{MLN86,Seifert05,Seifert07,Seifert08}. For stochastic systems,
the central problem is around the extension of the Second Law, which
originally describes the fundamental limitation on possible
transitions between equilibrium states. And recently, it has been
studied from the trajectory point of view, which stimulated the
rapid emergence of so-called fluctuation theorems
\cite{Seifert08,Jar2008}. Most of them are merely valid for specific
processes, such as steady states or transitions between steady
states; while the newly developed integral fluctuation theorems for
total entropy production and housekeeping heat could hold for
arbitrary time-dependent systems \cite{Seifert05,Speck05}.

These fluctuation theorems require the definitions of thermodynamic
functionals along the trajectory, and most of the have already done.
Thus the main purpose of the present article is to investigate the
thermodynamic laws within them, especially the two quantitative
forms of the Second Law.

Based on the stochastic processes, we put forward a rather
relatively integral theory of nonequilibrium thermodynamics, which
may be more definite and convincing but more restrictive than the
previous phenomenological frameworks. In addition, it would be
interesting to test experimentally all the quantities and relations,
especially in nonharmonic time-dependent potentials, where one does
not expect Gaussian distributions.

\begin{acknowledgments}
The author is grateful to Prof. Min Qian in Peking University and
Prof. Hong Qian in University of Washington for helpful discussions.
\end{acknowledgments}

\end{document}